\documentclass[aps, twocolumn, pra, showpacs, superscriptaddress]{revtex4}

\usepackage{graphicx}
\usepackage{dcolumn}
\usepackage{bm}
\usepackage{amsmath}

\begin{document}

\title{Preparation of stable excited states in an optical lattice via sudden quantum
quench}
\author{Li Wang}
\affiliation{Institute of Physics, Chinese Academy of Sciences, Beijing 100190, China}
\author{Yajiang Hao}
\affiliation{Department of Physics, University of Science and
Technology Beijing, Beijing 100083, China}
\author{Shu Chen}
\email{schen@aphy.iphy.ac.cn}
\affiliation{Institute of Physics, Chinese Academy of
Sciences, Beijing 100190, China}

\date{\today }
\begin{abstract}
We study how stable excited many-body states of the Bose-Hubbard
model, including both the gas-like state for strongly attractive
bosons and bound cluster state for repulsive bosons, can be produced
with cold bosonic atoms in an one-dimensional optical lattice.
Starting from the initial ground states of strongly interacting
bosonic systems, we can achieve stable excited states of the systems
with opposite interaction strength by suddenly switching the
interaction to the opposite limit. By exactly solving dynamics of
the Bose-Hubbard model, we demonstrate that the produced excited
state can be a very stable dynamic state. This allows the
experimental study of excited state properties of ultracold atoms
system in optical lattices.
\end{abstract}

 \maketitle

\section{Introduction}
Recent experiments with ultracold atoms have
offered exciting opportunities to study quantum many-body physics in
a highly controlled manner \cite {Review,Greiner,Kinoshita,Paredes}.
Beyond simulating the ground state (GS) properties of various
many-body systems, the uniqueness of cold atomic system, such as the
low dissipation rate and the Feshbach resonance technique, has led
to the experimental realization of stable excited states, as
demonstrated by recent observation of a one-dimensional (1D) stable
excited state called super-Tonks-Girardeau (STG) gas \cite{Haller}
and repulsively bound atom pairs \cite{Winkler}. In general, a
stable excited state is hard to be realized in traditional solid
state systems since a pure excited state is not stable due to the
energy dissipation between the system and the environment. The STG
gas provides one of the counterintuitive examples realized in cold
atom systems with no analog in solid state systems. It describes the
lowest gas-like phase of the attractive Bose gas
\cite{Astrakharchik1,Batchelor}, which is however a highly excited
state against its cluster-like GS. The stability of the STG gas
could be understood from the dynamics of the 1D integrable Bose gas
\cite{Chen}. Another example, i.e., the repulsively bound atom
pairs, is also counterintuitive at first glance since two atoms with
strongly repulsive on-site interaction usually repel each other.
Theoretical studies have revealed the existence of exotic
repulsively bound pairs in optical lattices
\cite{Wang,petrosyan,Song,Molmer}.

The experimental realization of stable excited states provides a promising new area for
searching novel quantum states in cold atom systems \cite{Kantian,Rosch}. However, the
diversity and mechanism of realizing stable excited many-body states are still not well
understood. Questions arise whether the STG gas can also be realized in a lattice
system as the Tonks-Girardeau (TG) gas \cite{Paredes} and whether the two independent
experiments \cite{Haller,Winkler} can be understood in the same theoretical framework.
In this work, we study how to prepare specific stable excited states, including both
the gas-like excited state in attractive regime and repulsively bound cluster of atoms,
for a Bose gas in a 1D optical lattice described by the basic Bose-Hubbard model (BHM).
By exactly solving quench dynamics problem of BHM, we first show that a stable gas-like
phase for the Bose gas with strongly attractive on-site interaction can be realized in
the 1D optical lattice by suddenly switching interactions from the strongly repulsive
regime to the attractive regime. Such an excited gas-like state avoids collapsing to
the atom-cluster GS even under very strongly attractive interaction and could be viewed
as a realization of STG gas in optical lattices. Furthermore, we show that a stable
repulsively bound cluster state composed of $N$-atoms can be also realized by sudden
switch of the interactions from the attractive side to the repulsive side. The
existence of the repulsively bound cluster in optical lattice is closely related to the
phenomenon of repulsively bound pairs. Our study suggests that the two seemingly
unconnected phenomena \cite{Haller,Winkler} could be understood in a unified
theoretical framework, i.e., they could be explained by calculating the sudden quench
dynamics of the BHM with different initial states.

\section{System and scheme}
We consider the system of ultracold bosonic
atoms in a 1D deep optical lattice, which can be described by the
BHM \cite{Jaksch,Fisher}
\begin{equation}
\hat{H}=-J\sum_{i } \left(\hat{b}_i^{\dagger }\hat{b}_{i+1}+
\hat{b}_{i+1}^{\dagger }\hat{b}_{i} \right)+ \frac
12U\sum_i\hat{n}_i(\hat{n}_i-1),  \label{BHH}
\end{equation}
where $\hat{b}_i^{\dagger }$ is the creation operator of bosons at
the $i$th site, $ J $ and $U$ denote the hopping strength and
on-site interaction, respectively. The ratio $U/J$ can be tuned by
varying the depth of the optical lattice and using Feshbach
resonance \cite {Jaksch}. For convenience, we set $J=1$ as the
energy scale. Despite the BHM having been studied by various methods
\cite{Fisher,Stoof,Roth}, the model is generally not exactly
solvable by analytical method \cite{Haldane}.
As the GS properties have been extensively studied, the properties of excited state and
related non-equilibrium physics based on the Hubbard model have recently attracted lots
of attentions \cite{Kantian,Rosch,Kollath,Roux,Polkovnikov,Haque,Heidrich-Meisner}. In
this work, we propose a scheme of preparing stable highly excited states of the BHM via
a sudden switch of interactions from the strongly repulsive regime to the attractive
regime and vice versa. Similar kind of quench has previously been used to create
metastable states in other systems \cite{Haller,Chen,Kantian}. We shall demonstrate how
an initially prepared GS translates to a highly excited state by solving the quantum
dynamics of BHM.

Suppose that the initial state $\left|\Psi_{ini}(t=0)\right
> = \left|\psi_0(U_0) \right >$ is prepared in the strongly interacting regime with
either $U_0>0$ or $U_0<0$, after a sudden switch to the opposite regime with
interaction strength $U$, the wave-function $\left| \Psi(t)\right>=e^{-iH(U)t}\left|
\Psi_{ini}(U_0)\right>$ can be calculated via
\begin{equation}
\left| \Psi(t)\right>=\sum_{n} e^{-iE_n t} c_n \left| \psi_n(U)\right>, \label{psit}
\end{equation}
where $c_n=\left< \psi_n(U)\right| \left. \psi_0(U_0)\right>$ with
$\left|\psi_n (U)\right>$ representing the $n$-th eigenstate of the
BHM with on-site interaction $U$. It is straightforward that
$|c_n|^2$ is the transition probability from the initial state to
the $n$-th eigenstate of $H(U)$. To study the quench dynamics of the
BHM, we shall scrutinize the full spectra and eigenstates of the BHM
both analytically and numerically.

\section{Bethe-ansatz solution for two-particle BHM}
We begin with the
two-particle problem of the BHM which is exactly solvable with the
aid of Bethe-ansatz (BA) method. Although the two-particle problem
is quite simple, its analytical result can provide us quite
instructive understanding to many-particle systems as we shall
discuss later. The BA wavefunction takes the form of $ |\Psi \rangle
= \sum_{x_1,x_2} \Psi(x_1,x_2) |x_1,x_2 \rangle$ with
\begin{equation}
\Psi(x_1, x_2) = A_{12} e^{i (k_{1}x_{1} + k_{2}x_{2})}+A_{21}e^{i
(k_{2}x_{1} + k_{1}x_{2})}   \label{ansatz}
\end{equation}
defined in the domain $x_{1} \leq x_{2}$. The wavefunction in the
other region $x_{2} \leq x_{1}$ can be obtained by the symmetry of
wavefunction.
Explicitly, the coefficients are given by $A_P=(-1)^P(\sin
k_{p_2}-\sin k_{p_1}-iU/2)$, where $P=\{p_1,p_2\}$ is one of the
permutations of $1,2$ and $(-1)^P=1$ or $-1$ for even or odd
permutation. Under periodic boundary condition, the quasimomenta
$k_j$ ($j=1,2$) fulfill the Bethe-ansatz equations (BAE)
\begin{eqnarray*}
\exp \left( ik_jL\right) &=&\frac{\sin k_l-\sin k_j-iU/2}{\sin
k_l-\sin k_j+iU/2}.
\end{eqnarray*}
The total momenta of the system is given by $K = k_1+k_2$ and the
eigenenergies are given by $ E=-2 \sum_{i=1}^N \cos k_i $.
\begin{figure}[tbp]
\includegraphics[width=9.5cm]{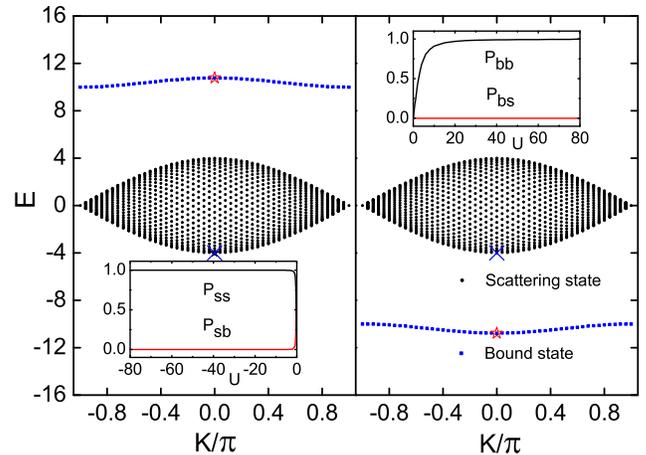}
\caption{(Color online) Full spectra for the two-particle BHM with
$L=50$, $U=10$ (left) and $U=-10$ (right) in $K$-space.  Left inset:
The transition probability from the initial scattering GS with
$U_0=80$ to the lowest scattering state ($P_{ss}$) and the dimer GS
($P_{sb}$) for various attractive $U$. Right inset: The transition
probability from the initial attractively dimer GS with $U_0=-80$ to
the repulsively dimer state ($P_{bb}$) and the scattering GS
($P_{bs}$) for various repulsive $U$.} \label{FigEnergy}
\end{figure}

The complete solutions of the BAE are composed of $C_L^2$ real
solutions (scattering states) and $L$ complex solutions (bound
states). The former ones follow by solving the logarithm of the
above BAE:
\begin{eqnarray}
k_jL &=&2\pi I_j+2\arctan \frac{\sin k_l-\sin k_j}{U/2}
\label{Scattering state}
\end{eqnarray}
with $I_j=-(L-1)/2,...,(L-1)/2$ and $I_1<I_2$. While the bound
states correspond to the string solution with the form of $k_1
=k+i\Lambda$ and $k_2=k-i\Lambda$, where $k$ and $\Lambda$ are real.
In this case the total momenta and eigenenergy take the form of
$K=2k$ and $E=-4J\cos k\cosh \Lambda$, respectively. In terms of $k$
and $\Lambda$ the original BAE transform into finding root of
\begin{equation}
\cos \left( kL\right) \exp \left( \Lambda L\right) =\frac{2\cos
k\sinh \Lambda -U/2}{2\cos k\sinh \Lambda +U/2},  \label{Bound
state}
\end{equation}
where $k=I_j\pi /L$ with $I_j=0,\pm 1,\pm 2,...,\pm L/2-1.$ Besides
these $L-1$ bound states the $L$th one corresponds to $K_L=\pi$ and
$E_L=U/J$. We obtain all eigenstates by solving (\ref{Scattering
state}) and (\ref{Bound state}). Full spectra for example systems
with $U=\pm 10$ and $L=50$ are displayed in Fig. \ref{FigEnergy}.

Taking advantage of analytical results for the two-particle case, we can calculate the
transition probabilities exactly. First, we consider the case in which the initial
state is prepared as the GS of the BHM in the repulsive side. After suddenly switching
the on-site interaction to the attractive regime, we calculate the overlap of the
initial wavefunction with the eigenstates of the attractive BHM. In the left inset, we
show the transition probability from an initial GS (marked by the cross in the left
Fig.1) with $U_0= 80$, to the lowest scattering state (marked by the cross in the right
Fig.1) with different values $U<0$ . Here the transition probability from the initial
scattering GS to the final lowest scattering state is denoted by $P_{ss}$ which is very
close to one after switching to the strongly attractive regime. On the other hand, the
probability for dynamically falling into the attractively bound state (denoted by
$P_{sb}$) is almost zero. In the right inset, it is shown that after switching to the
strongly repulsive regime, the initial attractively pair state (marked by the star in
the right Fig.1) is transformed to the repulsively bound state (marked by the star in
the left Fig.1) with the transition probability close to 1. Here the symbols $P_{bs}$
and $P_{bb}$ represent the transition probabilities from the initial bound state to
final scattering state and the initial bound state to final bound state, respectively.

\section{Many-particle system}
For the many-particle system with $N \geq 3$, the BHM is no longer
exactly solvable by the BA method which can not properly treat the
multi-occupation case \cite{Haldane}. Nevertheless, for a finite
size system, we can resort to the full exact diagonalization (ED)
method to calculate the full energy spectra and eigenstates.
Consequently the transition probabilities from the initial GS to
arbitrary final states are straightforward to be calculated. In
general, it is a formidable task to get the full spectra of a large
system as the basis dimension of a $N$-particle BHM with size $L$ is
given by $D=\left( N+L-1\right) !/[N!\left( L-1\right) !]$.

\begin{figure}[tbp]
\includegraphics[width=9.5cm]{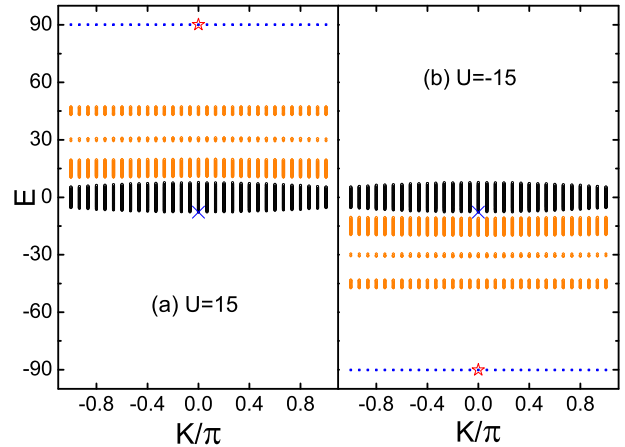}
\caption{(Color online) Full spectra of energies vs momentum for the
BHM with $N=4$, $L=30$ and $U=\pm 15$. The symbols of cross
represent the lowest scattering state in the bottom band (GS) for
the repulsive $U$ (left) and the lowest scattering state in the top
band for the attractive $U$ (right), whereas the symbols of star
represent the highest repulsively bound cluster state (left) and the
lowest attractively bound cluster state (right).} \label{4boson}
\end{figure}

Despite the full spectra becoming very complicated as the particle
number increases, we can still find some common characteristics of
the spectra for systems with different sizes. When $|U| \gg J$, the
spectra is split into a series of separated bands. For the repulsive
case, the lowest band is a scattering continuum of $N$
asymptotically free particles, whereas the top band is a narrow band
formed by the $N$-particle repulsively bound state. Between the top
and bottom bands, there exists a series of scattering continuums
formed by bound cluster states or formed by bound state and free
particle. To give a concrete example which may guide us to
understand the structure of the spectra, we display the
energy-momentum spectra for a system with $ N=4$, $L=30$ and $U=\pm
15$. As shown in Fig.\ref{4boson}a, the bound states on the top band
have the energies about $ 6 U $ , and the lowest band is a
scattering continuum of 4 free particles. In between, the energies
of three separated bands are approximately given by $3 U$, $2U$ and
$U$, which suggests that these bands correspond to, respectively,
the scattering continuum of a trimer and a free particle, two
dimers, and a dimer and two free particles.
The spectra for system with attractive interactions has a similar
structure in reverse order. The zero-momentum $N$-particle
attractively bound state is the GS, whereas the scattering continuum
of $N$ free particles is on the top of the spectra corresponding to
highly excited states of the attractive system. We note that these
separated bands are no longer discernable when the interaction
strength is comparable to the band width.

\begin{figure}[tbp]
\includegraphics[width=9.0cm]{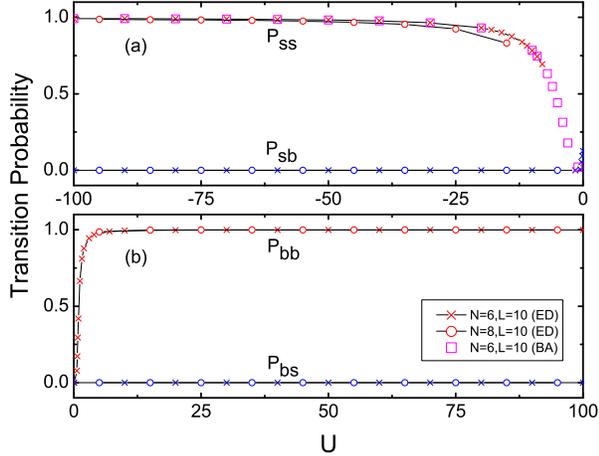}
\caption{(Color Online) (a) The transition probability from the
initial GS of the repulsive system with $U_0=100$ to the lowest
gas-like state ($P_{ss}$) and the cluster GS ($P_{sb}$) after the
switch to attractive side. (b) The transition probability from the
initial GS of the same system with $U_0=-100$ to the GS ($P_{bs}$)
and the repulsively cluster state ($P_{bb}$) for various repulsive
$U$. } \label{overlap}
\end{figure}

For a sudden switch of the interaction from the strongly repulsive
regime to the attractive side, we evaluate the transition
probabilities from the initial repulsively GS to the final states.
As displayed in Fig.\ref{overlap}a, the transition probability to
the lowest excited state in the top scattering band is very close to
1 in the strongly attractive regime, whereas the probability for
falling into the attractively cluster GS is almost completely
suppressed. Since the transition rate to the scattering phase is
very close to $1$ in the strongly interacting regime, we expect that
one can prepare such a highly excited state experimentally through
switching the interaction from strong repulsion into strong
attraction following the same way in the experiment of the STG gas.
Actually, the stable excited scattering state prepared in this way
can be viewed as a realization of STG gas in optical lattices. The
gas-like excited state is no longer stable and decays quickly if the
system enters to the weakly interacting regime.

\begin{figure}[tbp]
\includegraphics[width=8.0cm]{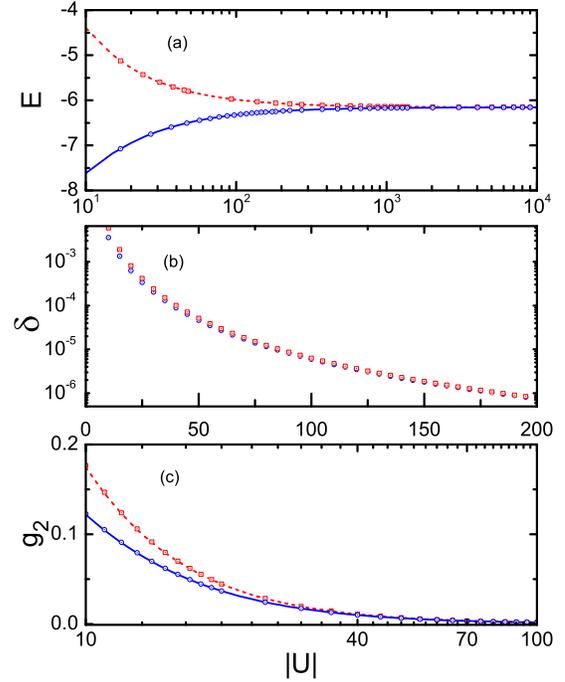}\newline
\caption{(Color Online) (a) $E_{TG}$ and $E_{STG}$ vs $|U|$ for system with $N=6$ and
$L=10$. Dash and solid lines are obtained by solving BAE whereas the squares (STG) and
dots (TG) are obtained by ED. (b) $\delta$ vs $|U|$ with
$\delta(U)=|E_{ED}(U)-E_{BA}(U)|/|E_{ED}(U)|$. Comparison between the BA solutions and
ED results suggests that the BA solutions can give very good results consist with the
ED results in the strongly interacting regime. (c) The local two-particle correlation
function $g_2 (U) =
\partial E(U) /\partial U$ vs interaction. } \label{g2}
\end{figure}

We can understand the stability of the lowest scattering state in
the strongly attractive limit from the analytical BA solution. We
note that the BAE solutions for the repulsive case and that for the
STG gas correspond to the same set of $\{I_j\}$ according to
(\ref{Scattering state}). In the limit $|U| \rightarrow \infty$, the
solutions given by $k_j = I_j 2 \pi/L$ are exactly the same, which
means that the repulsive GS state and the lowest scattering state in
the attractive side are identical in the infinitely interacting
limit. In the strongly interacting regime with $|U| \gg 1$, the
quasi-momentum distributions for the repulsive TG gas and the STG
gas approach the free fermion distribution from different sides, and
consequently the overlap between the repulsive GS and the STG state
is close to 1, i.e., $|\langle \psi_{STG}(-U)| \psi_0(U)\rangle|
\rightarrow 1$ as $|U| \rightarrow \infty$. The above exact
discussion for 2-particle system can be directly extended to the
many-particle systems. This is based on the observation that, in the
strongly interacting regime, the extended BA solutions
(\ref{BAE_manybody}) can be used to describe the properties of
scattering states very precisely for both the repulsive and
attractive cases, although the solution is not an exactly analytical
solution for the many-particle system in the rigorously integrable
meaning \cite{Haldane}. Here the many-body BA wavefunction takes the
form of $\Psi \left(
x_1,...,x_N\right)=\sum_PA_Pe^{i\sum\nolimits_jk_{p_j}x_j} $, where
the coefficients $A_P=(-1)^P\prod\nolimits_{j<l}^N(\sin k_{p_l}-\sin
k_{p_j}-iU/2)$, $P=\{p_1,p_2,...,p_N\}$ is one of the permutations
of $1,...,N$, and $\sum_P$ is the sum of all permutations. The
quasimomenta are determined by the BAE
\begin{equation}
k_jL=2\pi I_j+2\sum\nolimits_{l=1}^N\arctan \frac{\sin k_l-\sin
k_j}{U/2}. \label{BAE_manybody}
\end{equation}
Both the GS solution for $U>0$ and the STG solution for $U<0$ are determined by the
same set of $\{I_j\}=\{-(N-1)/2,...,(N+1)/2\}$. To confirm that, we calculate and
compare the GS energy for the $U>0$ case and energy of the lowest scattering state for
the $U<0$ case by both the BA solution and ED (see Fig.\ref{g2}a). The energies
obtained by the two methods agree very well with, for example,
$\delta=|E_{ED}-E_{BA}|/|E_{ED}| < 10^{-5}$ for $|U|=100$ as shown in the
Fig.\ref{g2}b. As $|U|\rightarrow \infty$, we have $E_{TG}=E_{STG}$. Such a gas-like
state shares similar characteristics with its continuum correspondence. For example,
the STG phase exhibits stronger local correlation than the repulsive Bose gas as
displayed in the Fig.\ref{g2}c.

Finally, we turn to the case in which the initial state is chosen to
be the GS of the BHM in the strongly attractive interaction limit,
i.e., the cluster-type bound state with all the atoms tending to
stay together due to the strongly on-site attractive interaction.
After the on-site interaction is suddenly switched to the strongly
repulsive regime, we also find that the cluster-type bound state is
stable and the initial system translates to the repulsively cluster
state as marked by star in Fig.\ref{4boson}a. In Fig.\ref{overlap}b,
we show the transition probabilities from the initial attractively
bound state to the repulsively bound state and the scattering GS in
the repulsive side. We find that the transition probability to the
repulsively bound state is close to 1 whereas the transition to the
scattering GS is almost completely suppressed in a wide range of
interaction.

\section{Summary}
In summary, we have studied the transition from the GS of strongly repulsive or
attractive bosons to the gas-like highly excited state of attractive bosons or the
repulsively bound state through a switch of interaction. By calculating the transition
probabilities, we have shown that the gas-like excited state and repulsively bound
state are stable in the strongly interacting regime and thus are possible to be
observed with cold atoms in optical lattices.

\begin{acknowledgments}
This work is supported by the NSF of China under Grant Nos. 10974234 and 10821403, 973
grant No. 2010CB922904 and National Program for Basic Research of MOST.
\end{acknowledgments}

\end{document}